\documentclass{aastex}

\newcommand{\emaila}{ E-mail: mfullana@mat.upv.es}
\newcommand{\emailb}{ E-mail: aalfonsofaus@yahoo.es}

\begin{document}

\title{Quantization of the Universe as a Black Hole
}

\shorttitle{Quantization of the Universe as a Black Hole}
\shortauthors{M. J. Fullana i Alfonso and A. Alfonso-Faus}

\author{M\`arius Josep Fullana i Alfonso\altaffilmark{1}} 
\affil{Institut de Matem\`atica Multidisciplin\`aria, \\
Universitat Polit\`ecnica de Val\`encia, \\
Cam\'{\i} de Vera, 
Val\`encia, 46022, Spain.\\
\emaila}
\and
\author{Antonio Alfonso-Faus\altaffilmark{2}}
\affil{Escuela de Ingenier\'{\i}a Aeron\'autica y del Espacio, \\
Plaza del Cardenal Cisneros, 3,
Madrid, 28040, Spain.\\
\emailb \\ 
. \\
Accepted for publication in Astrophysics \& Space Science \\
in October 25th 2011}

\begin{abstract}
It has been shown that black holes can be quantized by using Bohr's idea of quantizing the motion of an electron inside the atom. 
We apply these ideas to the universe as a whole. 
This approach reinforces the suggestion that it may be a way to unify gravity with quantum theory.
\end{abstract}


\section{Introduction}	

He and Ma (\citeyear{HeMa}) have shown that it is possible to consider black holes as quantized objects. Our approach here is to apply this idea to the universe. This is in line with the previous work that considers the universe to be a black hole  (\citeauthor{AAF1} \citeyear{AAF1}). We find here the equivalent quantum number for the universe and present the bit as the quantum of the gravitational potential, as already noted in \cite{AAF2}a. This number is close to $10^{122}$. We note that this big number is close to, if not the same, the worrying discrepancy that one has between the cosmological constant value, from cosmological observations, and the value obtained from the standard theory of particles (\citeauthor{AAF3} \citeyear{AAF3}). 
A likely explanation for this discrepancy has also been advanced by Santos (\citeyear{ES1}a, \citeyear{ES2}b and \citeyear{ES3}).
This strongly suggests that our universe can be considered to be an excited state of Planck's quantum black hole. Again, this is reinforced by the fact that the natural units, Planck's units, when multiplied by the big number $10^{61}$, give the well known physical properties of the universe, mass, length and time. Also the entropy of the universe, as a black hole, is linked to this quantum number being of the order of the square of it, nearly the same as the estimation made by Eagan and Lineweaver (\citeyear{EL}). Finally we refer to the consideration made by Lloyd 
(\citeyear{Llo}) in the sense of the universe as being similar to a quantum computer, where the big number $10^{120}$ plays the role of the maximum number of elementary operations made by the universe on $10^{90}$ bits  (\citeauthor{Llo} \citeyear{Llo}). Our estimate has gone further, the number of bits is found to be $10^{122}$ (\citeauthor{AAF4} \citeyear{AAF4}b).

\section{Quantization of the Universe as a Black Hole}

We reinforce the statement of He and Ma (\citeyear{HeMa}): {\it We anticipate that these ideas will lead to new understanding and perspective on gravity}.

The Schwarzschild radius $R = 2GM/c^2$ of a black hole may be combined with the Compton wavelength $\bar{\lambda} = \hbar /Mc$ to give (\citeauthor{HeMa}
\citeyear{HeMa}):

\begin{equation}
R \bar{\lambda} = 2 (l_p)^2
\label{eq.1}
\end{equation}

\noindent
where  $l_p$  is the Planck's length $l_p  = (Gh/c^3)^{1/2}$, $G$ the gravitational constant, $\hbar$ the Planck's constant and $c$ the speed of light. 
Relation (1) states that the geometric mean between $R$ and $\bar{\lambda}$ is of the order of Planck's length. For a black hole, mass and length are proportional. 
Since the Planck's units correspond to a quantum black hole; 
the generalization of the relation (1) gives as a result that all properties of a black hole, length $R$, mass $M$ and characteristic time $R/c$, are related in this way. 
It means that any black hole $(R,M)$ has a {\it conjugate} black hole given by the relation (1), with properties of mass $m$ and Schwarzschild radius $\bar{\lambda}$ such that (1) is satisfied. 

We have presented, and analyzed elsewhere (\citeauthor{AAF1} \citeyear{AAF1}), the universe as a black hole. There have also been arguments in favor of considering the universe, as a whole, 
to be a quantum object (\citeauthor{PFGD} \citeyear{PFGD}). The combination of these two ideas gives us a result that is related to expression (1): The universe behaves as a quantum black hole (\citeauthor{AAF1} \citeyear{AAF1}). All that is needed is to generalize Planck's constant. Applying (1) to the universe, with $R \approx 10^{28} cm$, mass $M \approx 10^{56} g$ and characteristic time $t_0 \approx  4.3 \ \times 10^{17} s$, we get the conjugate black hole:

\[
                           r \approx  \frac{2 (l_p)^2}{R}  \approx 5.2 \ 10^{-94} cm \]
\[
                            m \approx  \frac{2 (m_p)^2}{M} \approx  10^{-65} g                                      
\]
\begin{equation}
\label{eq.2} 
                           \tau \approx  \frac{2 (t_p)^2}{t_0}  \approx 1.35 \ 10^{-104} s \\
\end{equation}

\noindent
where $m_p = (\hbar c/G)^{1/2}$ is Planck's mass and   $t_p = (G \hbar /c^5)^{1/2}$ is Planck's time.

The mass $m \approx 10^{-65} g$ of the conjugate black hole of the universe has been identified with the quantum of the gravitational potential field
(\citeauthor{AAF2} \citeyear{AAF2}a) and the bit (\citeauthor{AAF4} \citeyear{AAF4}b). This is in line with the suggestion that this is a possible way to unify gravity with quantum theory. 
Besides, the information-entropy relation, based on the bit, 
the Padmanabhan (\citeyear{PMBa}a, \citeyear{PMBb}b and references therein)
proposal that gravity has an entropic or thermodynamic origin, 
and the Verlinde interpretation of gravity as an emerging entropic force
(\citeauthor{EPV} \citeyear{EPV}), gives us a hope in this direction.

The physical properties of the bit in (2) clearly imply a number of bits for the universe of $M/m \approx 10^{122}$. On the other hand the Hawking entropy for a black hole (\citeauthor{SWH} \citeyear{SWH}), (the Bekenstein limit, see 
\citeauthor{JB} \citeyear{JB}) is for the universe:

\begin{equation}
\label{eq.3}
S = \frac{4 \pi k}{\hbar c} \ G M^2 = \pi k \left ( \frac{R}{l_p}  \right)^2 \approx k 10^{122}
\end{equation}

Here we see that the entropy of the bit is just $k$, the Boltzmann's constant, and the information of the universe, as a black hole, is $\approx 10^{122}$ bits, its quantum number. 
All these results are in line with the results in \cite{HeMa} that we have
applied here to the universe. In particular the energy difference between two nearby states for the universe is 

\begin{equation}
\label{eq.4}
\Delta E \approx \frac{m_p c^2}{2 \ 10^{61}} \approx 10^{-65} \ \hbox{grams} \times c^2
\end{equation}

\noindent
which is the energy of one bit, the quantum black hole in (2). This means that each jump in the excitation of the universe implies precisely this energy. The excitation of the universe is given by the increase in its number of bits, one by one.

\section{Conclusions}

Our approach shows that the expansion of the universe implies the increase of entropy, and consequently the increase in its information content given by its quantum number. This increase can be explained as a process of sequential growing: 
this gives the chain of elements formed by a causal set of the initial black holes
(\citeauthor{SPG}\citeyear{SPG}) . This picture reinforces the idea of the universe as being a quantum computer, advanced by Lloyd (\citeyear{Llo}).

\end{document}